\documentclass[11pt,a4paper]{article}

\usepackage[a4paper,text={450pt,650pt},centering]{geometry}

\usepackage{hyperref}
\usepackage{mathrsfs}
\usepackage{amsmath,amssymb,dsfont,placeins}
\usepackage{verbatim}
\usepackage{stmaryrd,xcolor}
\usepackage{amsfonts,float,ulem}
\usepackage[pdftex]{graphicx}
\usepackage{bbm}
\usepackage{color}
\usepackage{lineno}

\renewcommand{\arraystretch}{0.8}

\font\tenfrak=eufm10  
\font\sevenfrak=eufm7 \font\fivefrak=eufm5
\newfam\frakfam
\textfont\frakfam=\tenfrak \scriptfont\frakfam=\sevenfrak
\scriptscriptfont\frakfam=\fivefrak

\newcommand\blfootnote[1]{%
  \begingroup
  \renewcommand\thefootnote{}\footnote{#1}%
  \addtocounter{footnote}{-1}%
  \endgroup
}

\begin{document}



\begin{titlepage}


\begin{centering}

\vspace{1in}

{\Large {\bf Parametrizing the Neutrino sector of the seesaw extension in tau decays}}${}^{\ast}$\blfootnote{${}^{*}$Presented at the 36th International Conference on High Energy Physics, 4-11 July 2012, Melbourne, Australia}

\vspace{.5in}
%
{\large D. Jur\v{c}iukonis${}^{a,1}$, T. Gajdosik${}^{b,2}$, A. Juodagalvis${}^{a,3}$ and T. Sabonis${}^{a,4}$}\\
\vspace{.3 in}
${}^{a}${\textit{Institute of Theoretical Physics and Astronomy, Vilnius University,\\Go\v{s}tauto g. 12, Vilnius 01108, Lithuania}}\\
${}^{b}${\textit{Vilnius University, Physics Faculty,\\Saul\.{e}tekio al. 9, Vilnius 10222, Lithuania}}\\

\footnotetext[1]{{\tt Darius.Jurciukonis@cern.ch,}\quad
${}^{2}${\tt Thomas.Gajdosik@cern.ch,}\quad ${}^{3}${\tt
Andrius.Juodagalvis@cern.ch}\\ \quad ${}^{4}${\tt
Tomas.Sabonis@tfai.vu.lt}\quad } \vspace{.5in}

\vspace{.1in}

\end{centering}

\begin{abstract}
The Standard Model includes neutrinos as massless particles, but
neutrino oscillations showed that neutrinos are not massless. A simple
extension of adding gauge singlet fermions to the particle spectrum
allows normal Yukawa mass terms for neutrinos. The smallness of the
neutrino masses can be well understood within the seesaw mechanism. We
analyse two cases of the minimal extension of the standard model when
one or two right-handed fields are added to the three left-handed
fields. A second Higgs doublet is included in our model. We calculate
the one-loop radiative corrections to the mass parameters which
produce mass terms for the neutral leptons. In both cases we
numerically analyse light neutrino masses as functions of the heavy
neutrinos masses. Parameters of the model are varied to find light
neutrino masses that are compatible with experimental data of solar
$\Delta m^2_\odot$ and atmospheric $\Delta m^2_\mathrm{atm}$ neutrino
oscillations for normal and inverted hierarchy.
\end{abstract}

\end{titlepage}


\section{The model}

We extend the Standard Model (SM) by adding a second Higgs doublet and 
right-handed neutrino fields. The Yukawa Lagrangian
of the leptons is expressed by

\begin{equation}
\mathcal{L}_\mathrm{Y} = - \sum_{k=1}^{2}\, 
\left( \Phi_k^\dagger \bar \ell_R \Gamma_k
+ \tilde \Phi_k^\dagger \bar \nu_R \Delta_k \right) D_L
+ \mathrm{H.c.}
\label{Yukawa}
\end{equation}
in a vector and matrix notation, where $\tilde \Phi_k = i \tau_2
\Phi_k^\ast$.  In expression (\ref{Yukawa}) the $\ell_R$, $\nu_R$, and
$D_L=(\nu_L\ \ell_L)^T$ are the vectors of the right-handed charged
leptons, of the right-handed neutrino singlets, and of the left-handed
lepton doublets, respectively, and $\Phi_{\alpha}, \alpha=1,2$ are the 
two Higgs doublets. The Yukawa coupling matrices $\Gamma_k$
are $n_L \times n_L$, while the $\Delta_k$ are $n_R \times n_L$.

In this model, spontaneous symmetry breaking of the SM gauge group is
achieved by the vacuum expectation values
$\langle\Phi_k\rangle_{\mathrm{vac}} =
\left(\renewcommand{\arraystretch}{0.2} \begin{array}{c} \scriptstyle
0 \\ \scriptstyle v_k/\sqrt{2} \end{array} \right)$, $k=1, 2$. By unitary
rotation of the Higgs doublets
we can achieve
$\langle\Phi^0_1\rangle_{\mathrm{vac}} = v/\sqrt{2} > 0$ and
$\langle\Phi^0_2\rangle_{\mathrm{vac}} = 0$ with $v \simeq 246$~GeV.
The charged-lepton mass matrix $M_\ell$ and the Dirac neutrino mass
matrix $M_D$ are

\begin{equation}
M_\ell = \frac{ v}{\sqrt{2}}\, \Gamma_1
\quad {\rm and} \quad
M_D = \frac{v}{\sqrt{2}}\, \Delta_1\, ,
\label{M_ell,M_D}
\end{equation}
with assumption that $M_\ell = {\rm diag} \left( m_e,
m_\mu, m_\tau \right)$. The mass terms for the neutrinos can be
written in a compact form with an $(n_L+n_R) \times (n_L+n_R)$
symmetric mass matrix

\begin{equation}\label{Mneutr}
M_{\nu} = 
\left(\renewcommand{\arraystretch}{0.8} \begin{array}{cc} 0 & M_D^T \\
M_D & \hat{M}_R \end{array} \right),
\end{equation}
where the hat indicates that $\hat{M}_R$ is a diagonal matrix. $M_{\nu}$ can be diagonalized as

\begin{equation}\label{Mtotal}
U^T M_{\nu}\, U = \hat m
= \mathrm{diag} \left( m_1, m_2, \ldots, m_{n_L+n_R} \right),
\end{equation}
where the $m_i$ are real and non-negative.  In order to implement the
seesaw mechanism \cite{seesaw,Schechter} we assume that the elements
of $M_D$ are of order $m_D$ and those of $M_R$ are of order $m_R$,
with $m_D \ll m_R$.  Then the neutrino masses $m_i$ with $i=1, 2,
\ldots, n_L$ are of order $m_D^2/m_R$, while those with $i = n_L+1,
\ldots, n_L+n_R$ are of order $m_R$.  It is useful to decompose the
$(n_L+n_R) \times (n_L+n_R)$ unitary matrix $U$ as $U =
\left(\renewcommand{\arraystretch}{0.2} \begin{array}{c} \scriptstyle
U_L \\ \scriptstyle U_R^\ast \end{array} \right)$, where the submatrix
$U_L$ is $n_L \times (n_L+n_R)$ and the submatrix $U_R$ is $n_R \times
(n_L+n_R)$ \cite{GN89,GL02}.  With these submatrices, the left- and
right-handed neutrinos are written as linear superpositions of the
$n_L+n_R$ physical Majorana neutrino fields $\chi_i$: $\nu_L = U_L P_L
\chi$ and $\nu_R = U_R P_R \chi$, where $P_L$ and $P_R$ are the
projectors of chirality.

We can express the couplings of the model in terms of mass
eigenfields, where three neutral particles are coupling to neutrinos.
Interaction of the $Z$ boson with the neutrinos is given by

\begin{equation}
\mathcal{L}_{\mathrm{nc}}^{(\nu)} = \frac{g}{4 c_w}\, Z_\mu
\bar \chi \gamma^\mu \left[ P_L \left( U_L^\dagger U_L \right)
- P_R \left( U_L^T U_L^\ast \right) \right] \chi\, ,
\label{Zinter}
\end{equation}
where  $g$ is the SU(2) gauge coupling constant and $c_w$ is the cosine of the Weinberg angle.

The full formalism for the scalar sector of the multi-Higgs-doublet SM is
given in Refs.~\cite{GN89,GL02}. The Yukawa couplings of the Higgs
bosons $H^0_b$ to the neutrinos are given by

\begin{equation}
\mathcal{L}_\mathrm{Y}^{(\nu)} \left( H^0 \right) = - \frac{1}{2 \sqrt{2}}\,
\sum_b H^0_b\, \bar \chi \left[
\left( U_R^\dagger \Delta_b U_L
+ U_L^T \Delta_b^T U_R^\ast \right) P_L
+ \left( U_L^\dagger \Delta_b^\dagger U_R
+ U_R^T \Delta_b^\ast U_L^\ast \right) P_R
\right] \chi \,,
\label{neutralYuk}
\end{equation}
with $\Delta_b = \sum_k b_k \Delta_k$, where $b$ are two-dimensional complex unit vectors. The neutral Goldstone boson $G_{b_Z}^0$ is given by the vector $b_Z$ with
$b_Z = \left(i,0 \right)$.

Once the one-loop corrections are taken into account
the neutral fermion mass matrix is given by

\begin{equation}\label{M1}
M^{(1)}_\nu = \left(\renewcommand{\arraystretch}{0.8} \begin{array}{cc}
\delta M_L & M_D^T+\delta M_D^T \\
M_D+\delta M_D    & \hat{M}_R+\delta M_R\end{array} \right)\approx
\left(\renewcommand{\arraystretch}{0.8} \begin{array}{cc}
\delta M_L & M_D^T \\
M_D   & \hat{M}_R\end{array} \right),
\end{equation}
where the $0_{3\times3}$ matrix appearing at tree level (\ref{Mneutr})
is replaced by the contribution $\delta M_L$. This correction is a
symmetric matrix, it dominates among all the sub-matrices of corrections.
Neglecting the sub-dominant pieces in (\ref{M1}), one-loop corrections to the neutrino masses originate via the self-energy function $\Sigma_L^S (0)=\Sigma_L^{S(Z)}(0)+\Sigma_L^{S(G^0)}(0)+\Sigma_L^{S(H^0)}(0)$, where the $\Sigma_L^{S(Z,G^0,H^0)}(0)$ functions arise from the
self-energy Feynman diagrams and are evaluated at zero external momentum squared. In the calculation of the self energies
the neutrino couplings 
are determined by 
eqs.\ (\ref{Zinter}) and (\ref{neutralYuk}).
Each diagram contains a divergent
piece but when summing up the three contributions the result turns out
to be finite.

The final expression for one-loop corrections is given by \cite{Grimus:2002nk} 

\begin{eqnarray}
\delta M_L &=&
\sum_{b} \frac{1}{32 \pi^2}\, \Delta_b^T U_R^\ast \hat m
\left(
    \frac{\hat {m}^2}{m_{H^0_b}^2}-\mathbbm{1}
  \right)^{-1}\hspace{-5pt}
  \ln\left(\frac{\hat {m}^2}{m_{H^0_b}^2}\right) U_R^\dagger \Delta_b \notag \\
&&+ \frac{3 g^2}{64 \pi^2 m_W^2}\, M_D^T U_R^\ast \hat m
\left(
    \frac{\hat {m}^2}{m_Z^2}-\mathbbm{1}
  \right)^{-1}\hspace{-5pt}
  \ln\left(\frac{\hat {m}^2}{m_Z^2}\right) U_R^\dagger M_D,
\label{corrections}
\end{eqnarray}
where the sum $\sum_b$ runs over all neutral physical Higgses $H^0_b$. 

\section{Case $n_R=1$}
\label{3x1}

First we consider the most minimal extension of the standard model by adding only one right-handed field $\nu_R$ to the three left-handed fields contained in $\nu_L$.

We parametrize $\Delta_1 = \frac{\sqrt{2}\, m_D}{v}\,\vec{a_1}^T$ and $\Delta_2 = \frac{\sqrt{2}\, m_D}{v}\,\vec{a_2}^T$ with $|\vec{a_1}|=1$ and $|\vec{a_2}|=1$. The symmetric mass matrix $M_{\nu}$ (\ref{Mneutr}) in block form is diagonalized by
\begin{equation}\label{Mneutr1}
U^{T}M_{\nu}U = 
U^{T}\left(\renewcommand{\arraystretch}{0.8} \begin{array}{cc} 0 & m_D \vec{a_1} \\
m_D \vec{a_1}^T & \hat{M}_R \end{array} \right)U=\left(\renewcommand{\arraystretch}{0.8} \begin{array}{cc} \hat{M}_l & 0 \\
0 & \hat{M}_h \end{array} \right).
\end{equation}
The non zero masses in $\hat{M}_l$ and $\hat{M}_h$ are determined analytically by finding eigenvalues of the hermitian matrix $M_{\nu}M^{\dagger}_{\nu}$. These eigenvalues are the squares of the masses of the neutrinos $\hat{M}_l=\text{diag}(0,0,m_l)$ and $\hat{M}_h=m_h$. The solutions $m^2_D=m_hm_l$ and $m^2_R=(m_h-m_l)^2 \approx m^2_h$ correspond to the seesaw mechanism.

We can construct the diagonalization matrix $U$ for the tree level from two diagonal matrices of phases and three rotation matrices $U_{\text{tree}}=U_{\phi}(\phi_i)U_{12}(\alpha_1)U_{23}(\alpha_2)U_{34}(\beta)U_i$, where the angle $\beta$ is determined by the masses $m_l$ and $m_h$. The values of $\phi_i$ and $\alpha_i$ can be chosen to cover variations in $M_D$.

For calculation of radiative corrections we use the following set of
orthogonal complex vectors: $b_Z = (i,0)$, $b_1 = (1,0)$, $b_2 =
(0,i)$ and $b_3 = (0,1)$. Diagonalization of the mass matrix after
calculation of one-loop corrections is performed with a unitary matrix
$U_{\text{loop}}=U_{\text{egv}}U_{\varphi}(\varphi_1,\varphi_2,\varphi_3)$,
where $U_{\text{egv}}$ is an eigenmatrix of
$M^{(1)}_{\nu}M^{(1)\dagger}_{\nu}$ and $U_{\varphi}$ is a phase
matrix. The second light neutrino obtains its mass from radiative
corrections. The third light neutrino remains massless.

\begin{figure}[H]
\begin{center}
\includegraphics[scale=0.53]{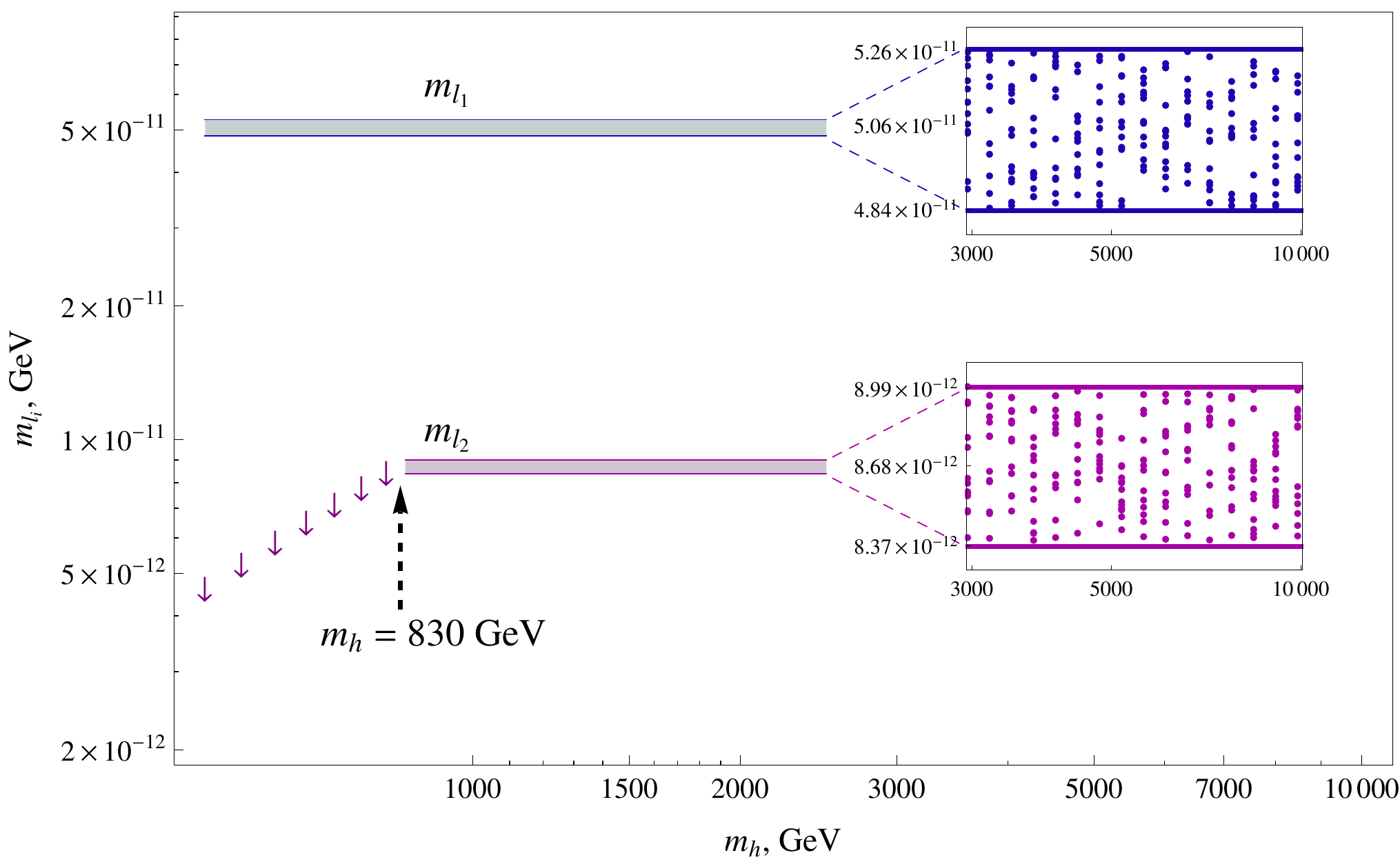}
\end{center}
\caption{Calculated masses of two light neutrinos as a function of the
  heavy neutrino mass $m_h$. The mass of the third light neutrino is
  zero, when $n_R=1$. Solid lines show the boundaries of the allowed
  neutrino mass ranges when the model parameters are constrained by
  the experimental data on neutrino oscillations. The purple arrows
  indicate the values of $m_{l_2}$ which do not satisfy
  the allowed experimental neutrino mass ranges. Due to the scale, the
  allowed $m_{l_1}$ and $m_{l_2}$ values form a band, their scattered
  values are shown separately in the right
  plots.}
\label{picture1}
\end{figure}

It is possible to estimate masses of the light neutrinos from experimental data of solar and atmospheric neutrino oscillations \cite{Gonzalez} assuming that the lightest $m_{l_3}=0$. Considering the normal ordering of the light neutrinos we receive $m_{l_1}=5.0\pm0.2\times 10^{-11}$~GeV and $m_{l_2}=8.7\pm0.3\times 10^{-12}$~GeV. Numerical analysis shows that we can reach those values for a heavy singlet with a mass bigger than 830~GeV, see Fig.~\ref{picture1}. 

\section{Case $n_R=2$}
\label{3x2}

If we add two singlet fields $\nu_R$ to the three left-handed fields
$\nu_L$, the radiative corrections give masses to all three light
neutrinos.

Now we parametrize $\Delta_1=\frac{\sqrt{2}}{v}\,\left(\renewcommand{\arraystretch}{0.2} \begin{array}{cc}\scriptstyle m_{D_2} \vec{a_1}^T \\ \scriptstyle m_{D_1} \vec{b_1}^T \end{array} \right)$ and $\Delta_2=\frac{\sqrt{2}}{v}\,\left(\renewcommand{\arraystretch}{0.2} \begin{array}{cc}\scriptstyle m_{D_2} \vec{a_2}^T \\ \scriptstyle m_{D_1} \vec{b_2}^T \end{array} \right)$
with $|\vec{a_1}|=1$, $|\vec{b_1}|=1$, $|\vec{a_2}|=1$ and $|\vec{b_2}|=1$. Diagonalizing the symmetric mass matrix $M_{\nu}$ (\ref{Mneutr}) in block form we write: 

\begin{equation}\label{Mneutr2}
U^{T}M_{\nu}U = 
U^{T}\left(\renewcommand{\arraystretch}{0.6} \begin{array}{cc} 0_{3 \times 3} & m_{D_2} \vec{a} \hspace{.2cm} m_{D_1} \vec{b}\\
\begin{array}{c} m_{D_2} \vec{a}^T \\ m_{D_1} \vec{b}^T \end{array} & \hat{M}_R \end{array} \right)U=\left( \begin{array}{cc} \hat{M}_l & 0 \\
0 & \hat{M}_h \end{array} \right).
\end{equation}
The non zero masses in $\hat{M}_l$ and $\hat{M}_h$ are determined by
the seesaw mechanism: $m^2_{D_i}\approx m_{h_i}m_{l_i}$ and $m^2_{R_i}
\approx m^2_{h_i}$, $i=1,2$. Here we use the ordering of masses $m_1>m_2>m_3$. 
The third light neutrino is massless at tree level.

The diagonalization matrix for tree level $U_{\text{tree}}=U_{12}(\alpha_1,\alpha_2)U_{\text{egv}}(\beta_i)U_{\phi}(\phi_i)$ is composed of a rotation matrix, an eigenmatrix of $U_{12}^T M_{\nu}M^{\dagger}_{\nu}U_{12}^\ast$ and a diagonal phase matrix, respectively.

For calculation of radiative corrections we use the same set of the
orthogonal complex vectors $b_i$ as in the first case. Diagonalization
of the mass matrix including the one-loop correction is performed with
a unitary matrix
$U_{\text{loop}}=U_{\text{egv}}U_{\varphi}(\varphi_i)$, where
$U_{\text{egv}}$ is the eigenmatrix of
$M^{(1)}_{\nu}M^{(1)\dagger}_{\nu}$ and $U_{\varphi}$ is a phase
matrix.

\begin{figure}[H]
\begin{center}
\includegraphics[scale=0.5]{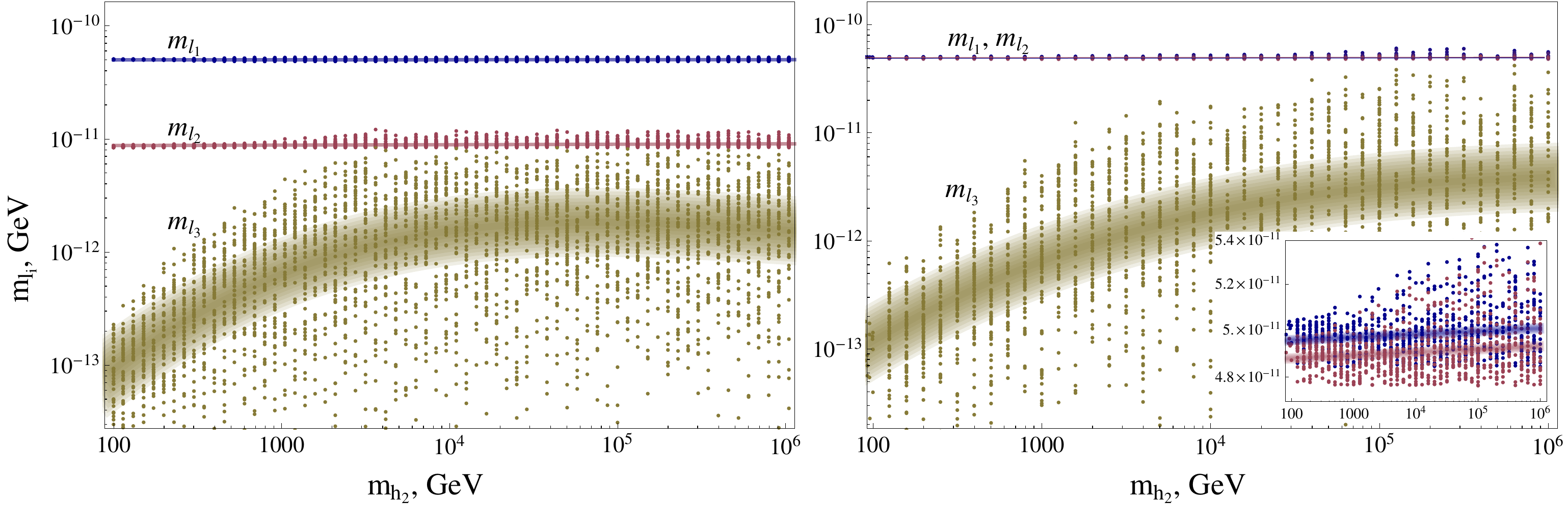}
\end{center}
\caption{The masses $m_{l_i}$ of the light neutrinos as functions of
  the heaviest right-handed neutrino mass $m_{h_1}$, for the case
  $n_R=2$. The value of the second heaviest right-handed neutrino mass is
  fixed at $m_{h_2}=100$~GeV. The plot on the left represents the normal
  hierarchy, and the plot on the right represents the inverted hierarchy of
  the light neutrinos. The wide solid lines indicate the area of the
  most frequent values of the scatter data.
  The nearly degenerate masses $m_{l_1}$ and $m_{l_2}$ are shown
  separately in the lower right plot for the case of the inverted hierarchy.
}
\label{picture2}
\end{figure}

In numerical calculations the model parameters as well as the derived
masses of the light neutrinos are obtained in several steps.  First,
the diagonal mass matrix for the tree level is constructed.  The lightest
neutrino is massless, and the masses of other two light neutrinos are
estimated from experimental data on solar and atmospheric neutrino
oscillations. The masses of the heavy neutrinos are input
parameters. This diagonal tree level matrix is used to constrain the parameters
$\alpha_i$ and $\phi_i$ that enter the tree-level mass matrix $M_\nu$
and its diagonalization matrix $U_{\text{tree}}$.
Then the diagonalization matrix is used to
evaluate one-loop corrections to the mass matrix. Diagonalization of
the corrected mass matrix yields masses for three light neutrinos. If the
calculated mass difference is compatible with the experimental
neutrino mass difference, the parameter set is kept. Otherwise,
another set of parameters is generated. Figure\ \ref{picture2} illustrates 
the obtained results. Both normal and inverted neutrino mass orderings are considered.

\section{Conclusions} \label{concl}

For the case $n_R=1$ we can match the differences of the calculated
light neutrino masses to $\Delta m^2_\odot$ and $\Delta
m^2_\mathrm{atm}$ with the mass of a heavy singlet larger than
830~GeV. Only normal ordering of neutrino masses is possible.

In the case $n_R=2$ we obtain three non vanishing masses of light
neutrinos for normal and inverted hierarchies. The numerical analysis
shows that the values of the light neutrino masses (especially of the
lightest mass) depend on the choice of the heavy neutrinos masses. The
radiative corrections generate the lightest neutrino mass and have a
big impact on the second lightest neutrino mass.

In future we plan to apply our parametrization to study the $\tau$
polarization coming from the decay of a $W$ boson in the data of the
CMS experiment at LHC and thus determine restrictions to the
parameters of the neutrino sector.

\bigskip \bigskip
\noindent \textbf{\large Acknowledgements} 

The authors thank Luis Lavoura for valuable discussions and suggestions. 
This work was supported by European Union Structural Funds project "Postdoctoral Fellowship Implementation in Lithuania".

\bigskip
\newcommand{\hepth}[1]{\href{http://arxiv.org/abs/hep-th/#1}{\tt hep-th/#1}}
\newcommand{\hepph}[1]{\href{http://arxiv.org/abs/hep-ph/#1}{\tt hep-ph/#1}}
\newcommand{\nuclth}[1]{\href{http://arxiv.org/abs/nucl-th/#1}{\tt hep-ph/#1}}
\newcommand{\arXiv}[1]{\href{http://arxiv.org/abs/#1}{\tt arXiv:#1}}

 \end{document}